\documentclass[a4paper,11pt]{article}
\usepackage{pos}

\newcommand{\mrm}[1]{\mathrm{#1}}
\newcommand{\mbr}[1]{\overline{\mathrm{#1}}}

\renewcommand{\logo}{\relax}

\title{Soft QCD theory}
\author*[a]{Torbj\"orn Sj\"ostrand}
\affiliation[a]{Department of Astronomy and Theoretical Physics,
  Lund University, \\
  S\"olvegatan 14A, SE-223 62 Lund, Sweden}

\emailAdd{torbjorn.sjostrand@thep.lu.se}
\abstract{Soft QCD is beyond perturbative control, and therefore
  phenomenological models have to be developed. These are implemented
  and combined within event generators. Typical aspects considered are
  multiparton interactions, colour reconnection, and hadronization.
  Of special interest are non-universal features of hadronization
  at the LHC, such as the recently observed enhancement of charm and
  bottom baryons, and the bottom baryon asymmetry, as well as the some
  years earlier observed strangeness enhancement. Several new production
  mechanisms, such as junction reconnection, ropes and shove, have been
  proposed to address some of these issues. Alternatively, an admixture of
  a quark--gluon plasma component also in pp collisions is introduced in
  the core--corona approach.}

\FullConference{%
  The Tenth Annual Conference on Large Hadron Collider Physics - LHCP2022\\
  16-20 May 2022\\
  online
}

\begin{document}
% Show LUTP number for arXiv submission
\begin{flushright}
LU TP 22--52\\
August 2022
\end{flushright}
\maketitle

\section{Introduction}

\begin{figure}[t]
\includegraphics[width=\textwidth]{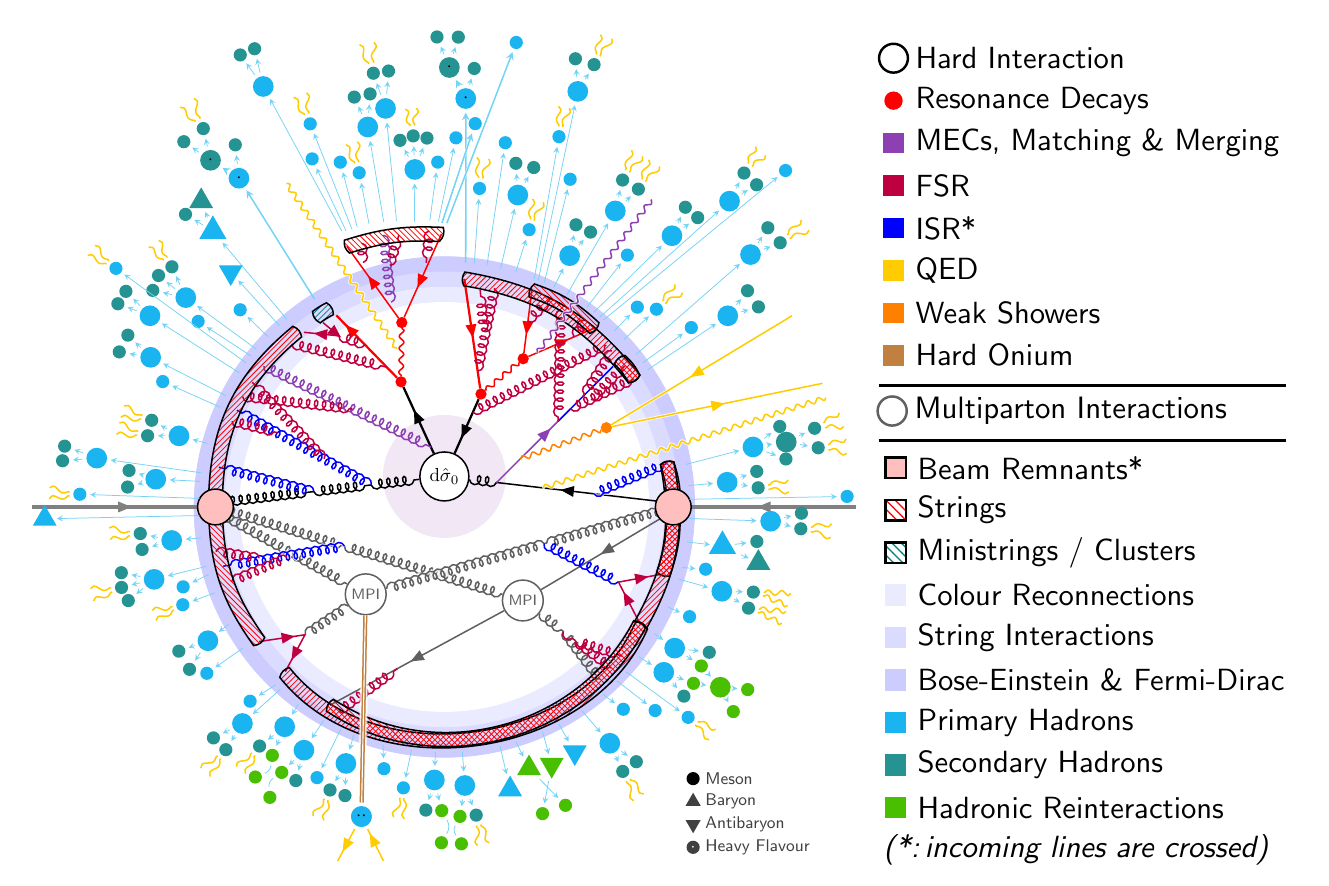}
\caption{The main components in the structure of an event,
illustrated for the case of top pair production, and approximately
ordered from hard to soft in the list.
Adapted from \cite{Bierlich:2022pfr}.}
\label{tsfig1}
\end{figure}

The structure of a typical pp collision is illustrated in Fig.~\ref{tsfig1},
split into the aspects we need to combine to obtain an overall description.

In the perturbative domain the hard interaction process defines the core,
calculated by the combination of matrix elements and parton distributions.
Decays of resonances such as W, Z, H and t also should be included in this
core. Higher-order matrix element corrections, on the one hand, and parton
showers, on the other, contribute to the emergence of multiparton topologies,
and can partly overlap. Special matching and merging procedures therefore
are devised to avoid doublecounting, and to provide a smooth transition
between the two descriptions. The showers can be subdivided into initial-
and final-state radiation, and may also include QED and weak emissions,
and even the perturbative emission of charmonium and bottomonium states.

Since hadrons are composite, multiple parton--parton interactions may occur
inside a single pp one. These straddle the perturbative--nonperturbative
borderline.

Central in the soft/nonperturbative phase, occurring at later invariant time
scales than the hard/perturbative one, is the formation of colour-confining
force fields, called strings or clusters, that subsequently fragment to
produce the primary set of hadrons. These fields are drawn between the
partons produced in the hard phase of the event, plus the beam-remnant
partons that passed through the interaction region unharmed. Before the
fragmentation step the fields may interact with each other, notably by
colour reconnection. After it, primary hadrons may decay over a wide range
of time scales. They may also rescatter against each other in the early
stages, when the hadronic density is high. Bose--Einstein and Fermi-Dirac
statistics can affect the production of identical particles during
the fragmentation, decay and rescattering stages.

Given the need to combine many disparate effects over a wide range of
scales, the most convenient approach is that of (Monte Carlo) event
generators \cite{Buckley:2011ms}. Here random numbers are used to emulate
the quantum mechanical uncertainty at each step of the construction of the
final state, and different code pieces can be called on to perform the
required range of tasks. Generators can be used to predict event properties,
for detector and trigger design, to correct data, for acceptance and smearing, 
and to interpret data in terms of underlying physics mechanisms.

The three most commonly used generators that combine hard and soft physics
are \textsc{Herwig} \cite{Bellm:2019zci}, \textsc{Pythia}
\cite{Bierlich:2022pfr} and \textsc{Sherpa} \cite{Sherpa:2019gpd}.
Generators for matrix elements and their interface to parton showers
notably include \textsc{MadGraph5\_aMC@NLO} \cite{Alwall:2014hca}
and the \textsc{PowHeg Box} \cite{Alioli:2010xd}. A wider range of
generators exists for applications to other fields, such as heavy ion
collisions or cosmic ray cascades, often modelling only soft-physics
aspects \cite{Campbell:2022qmc}. Of special note is \textsc{Epos}
\cite{Pierog:2013ria}, which spans the range between pp and heavy-ion
collisions.

\section{Multiparton interactions}

Multiparton interactions (MPIs) are essential to explain the bulk
properties of events, such as multiplicity distributions. Nevertheless
the underlying theory and the modelling thereof is still hotly debated
\cite{Bartalini:2018qje}. MPIs must extend down to a transverse momentum
$p_{\perp}$ of 2--3 GeV to generate the observed amount of activity in
events, a number that would seem to be in a perturbative regime, but
the lower turnoff must come from nonperturbative physics, such as the
screening of colour charges within the proton.

A similar dichotomy exists in the study of two hard subcollisions,
double parton scattering (DPS), a subset of the MPIs. The standard
formula for the cross section of simultaneously having processes
$A$ and $B$ in an event is
\[
  \sigma_{AB}^{\mrm{DPS}} = \frac{m}{2} \,
  \frac{\sigma_A \, \sigma_B}{\sigma_{\mrm{eff}}} \quad
  \mrm{where}~m = 1~\mrm{if}~A = B~\mrm{and}~m = 2~\mrm{else}.  
\]
Here $\sigma_A$ and $\sigma_B$ are perturbatively calculable, while
$\sigma_{\mrm{eff}}$ is a nonperturbative number. The latter 
is related to the nondiffractive cross section, of order 50~mb at LHC
energies, but then reduced when the impact-parameter profile is
considered. If protons are approximated by Gaussian distributions 
then $\sigma_{\mrm{eff}} \approx 25$~mb, and a picture with
``hot spots'' would give even lower numbers, i.e.\ more activity. 
Flavour and momentum conservation have opposite effects. 
Consider e.g.\ W$^-$W$^-$ production in the forward direction, where
one W but not both could benefit from a high-$x$ valence d quark.

\enlargethispage{0.5\baselineskip}
By now, several $\sigma_{\mrm{eff}}$ measurements have been done at
the LHC, but there is a wide spread of results, in the range
$5 - 20$~mb. It is not yet clear where this spread comes from.
Numbers at the lower end of the range also would suggest an overall
much larger MPI activity than what inclusive measurements do. An
interesting CMS study of 4-jet production \cite{CMS:2021lxi} shows
that more than a factor of 2 spread comes from using different
perturbative input, say leading vs.\ next-to-leading order. 
All data points also come with large error bars, that reflects
the difficult separation of signal from background, the uncertain
impact of more than two MPIs for the level of the underlying event,
and possibly more. In the future it will become important to analyze
several different processes under as identical conditions as possible,
to better understand the origin of the current $\sigma_{\mrm{eff}}$
spread.

\section{Colour reconnection and baryon enhancement}

The MPI activity leads to many colour fields --- strings --- being pulled
out in pp collisions, most of them running essentially parallel with each
other at low transverse momenta. Naively these strings would all be drawn
out to the beam remnants, i.e.\ to the colour ``hole'' left behind by
each parton that initiates an MPI.  In colour reconnection (CR)
scenarios the strings are drawn out less than expected.
One reason could be that the starting picture assumes infinitely many 
colours, such that each string piece is uniquely defined, while the
correct QCD picture provides ambiguities between identical colours.
Reassignments that lead to shorter strings, i.e.\ smaller invariant
masses,  are then likely to be dynamically preferred. The CR picture explains
e.g.\ why the average transverse momentum increases as a function of the
charged multiplicity. At LEP2 the hadronic final state of W$^+$W$^-$ is
best described with a $\sim$50\% CR rate \cite{ALEPH:2006bhb}, but
is still only $2.2 \sigma$ away from no-CR.

New CR models continue to appear. One such is the QCDCR one
\cite{Christiansen:2015yqa}. In it, regular
$\mrm{q}_1\mbr{q}_2 + \mrm{q}_3\mbr{q}_4 \to
\mrm{q}_1\mbr{q}_4 + \mrm{q}_3\mbr{q}_2$ reconnections are supplemented
by ones that produce junction topologies, i.e.\ where three color lines
come together in a \textsf{Y}-shaped configuration. In the
$\mrm{q}_1\mbr{q}_2 + \mrm{q}_3\mbr{q}_4$ example, if $\mrm{q}_1$ and
$\mrm{q}_3$ are nearby on the left side of the event, and $\mbr{q}_2$ and
$\mbr{q}_4$ ditto on the right one, the two long strings can collapse to
one in the central region, giving
\rotatebox[origin=c]{90}{\textsf{Y}}\rotatebox[origin=c]{270}{\textsf{Y}}
topologies.
Here the left-side junction will be associated with the production of
a baryon and the right-side antijunction with that of an antibaryon.
Thus an extra source of baryon production is introduced, and one that
becomes more important in high-multiplicity events, where more strings
run parallel and can reconnect this way.

Extending this approach to three parallel strings, it is possible to
obtain two disconnected junction systems,
$\mrm{q}_1\mbr{q}_2 + \mrm{q}_3\mbr{q}_4 + \mrm{q}_5\mbr{q}_6 \to 
\mrm{q}_1\mrm{q}_3\mrm{q}_5 + \mbr{q}_2\mbr{q}_4\mbr{q}_6$.
This latter mechanism has also been implemented in the \textsc{Herwig}
cluster fragmentation model \cite{Gieseke:2017clv}.

This approach has come into the limelight by the recent ALICE observation
of enhanced charm and bottom baryon production. Notably
the fraction of charm quarks that become $\Lambda_{\mrm{c}}^+$ is slightly
above 20\% \cite{ALICE:2021dhb}, about a factor 3 more than 
observed at LEP. This enhancement is concentrated to $p_{\perp}$ scales
below 10~GeV \cite{ALICE:2021rzj}, and above that the LEP numbers are
approached. The data is well
explained by the QCDCR model, where CR is most likely to occur in
the packed low-$p_{\perp}$ region, while high-$p_{\perp}$ jets mainly
evolve in a vacuum. Also some models based on quark recombination,
thermodynamics or quark--gluon plasma formation give comparable results.
A similar $p_{\perp}$-dependent behaviour has also been observed
for $\Lambda_{\mrm{b}}$ by LHCb \cite{LHCb:2019fns}.
One should add that the QCDCR model fails to
describe all the relevant distributions, and thus is not the ultimate
answer. It seems very likely, however, that baryon production by CR
will become a main staple in the further development of CR and
hadronization models. 

\section{Beam drag and forward physics}

In fixed-target experiments it has been known since long that the flavour
content of the beams influences the production of heavy hadrons. In $\pi^-$p
collisions, e.g., D$^-$ production increasingly dominates over D$^+$ 
one at higher momenta in the $\pi^-$ direction. This is easily understood
for the subprocess $\mbr{u}\mrm{u} \to \mbr{c}\mrm{c}$, where the
$\mbr{c}$ is colour-connected to the d in the $\pi^-$ remnant, while the
c is connected to the ud proton remnant. Then the drag of the colour
fields pulls $\mbr{c}$ forwards and c backwards \cite{Norrbin:2000zc}.
A low $\mbr{c}$d invariant mass even leads to a collapse directly into
a D$^-$.

Similar effects are predicted both for charm and bottom at the LHC,
leading to small asymmetries also att central rapidities. Asymmetries of
around 2\% have now been observed by LHCb, favouring $\Lambda_{\mrm{b}}$ 
over $\overline{\Lambda}_{\mrm{b}}$ \cite{LHCb:2021xyh}. The sign and
the slowly rising rapidity dependence of the effect
agrees with default \textsc{Pythia}, but is about a factor of 2 lower.
Interestingly, the QCDCR scenario agrees much better with data. Likely
the junction mechanism adds a symmetric source
of $\Lambda_{\mrm{b}}$ and $\overline{\Lambda}_{\mrm{b}}$ production that
dilutes the asymmetry present in the traditional baryon production.
Furthermore, in default \textsc{Pythia} the asymmetry is peaked at
small $p_{\perp}$, i.e.\ where the b/$\mbr{b}$ quark is closer to the
beam direction, while data and QCDCR show only a modest $p_{\perp}$
dependence. This is again consistent with a junction mechanism that
is most active at small $p_{\perp}$, and therefore gives the largest
dilution effect there.

More generally, our understanding of the forward region is still
limited. Traditionally this has been a region of special interest
mainly for cosmic-ray people, hence the LHCf experiment. Now
new-particle searches and neutrino studies in the forward direction
attracts increasing interest at the LHC. Comparisons with models have
shown significant disagreements in many cases, and studies are under way
to improve the situation. This may require a better modelling of beam
remnants and diffraction, among others.

\section{Hadronization}

The ALICE observation of an increasing strangeness fraction with
increasing multiplicity \cite{ALICE:2016fzo} was unexpected. The trend
smoothly attaches to results for heavy-ion runs, where the high
strangeness rate is coming from thermodynamics at the freezeout
temperature, when the quark--gluon plasma (QGP) turns into hadrons.
Traditional wisdom held that pp systems would not give sufficiently
large and long-lived regions for a QGP to form, but this is now cast
in doubt.

The simplest and most appealing solution is offered by core--corona
models, notably the \textsc{Epos} one. In it, initially all colour
fields are drawn out from all partonic interactions. But then, in
regions where these fields are very closely packed, it is assumed
that they together melt into a QGP state. A typical topology therefore
would have a dense QGP core surrounded by a normal-state dilute corona.
A low-multiplicity pp collision would be all corona, and thus have
the same particle composition as in e$^+$e$^-$, but with increasing
multiplicity the fraction of core increases, and thereby the relative
rate of strangeness production. In a nucleus--nucleus (AA) collision the
core region would dominate, except for the most peripheral collisions,
and the strangeness fraction saturate.

Nevertheless, it is interesting to consider scenarios that could avoid
the introduction of QGP in pp collisions, and maybe even offer alternative
explanations for (some of) the signals interpreted as proof of QGP
in AA. One such example is rope formation \cite{Bierlich:2014xba}, wherein
nearby strings can fuse to objects of a larger colour representation.
This gives a larger energy release when the ropes break, which translates
into a smaller strangeness suppression than in a single string. Higher
multiplicities means more strings and more possibilities for rope formation,
such that the rope scenario is in reasonable agreement with data. 
Another possible mechanism is shove, the repulsion of nearby strings,
which can give them a transverse motion inherited by the hadrons
produced from them, resulting in a flow-like behaviour
\cite{Bierlich:2016vgw}. While these and other ideas are very
interesting, it remains to put it all together in an combined framework
that gives a convincing picture at least of pp collisions
without any need for a QGP.

\section{Summary and outlook}

Before LHC, the concept of jet universality was commonly accepted.
That is, once the perturbative evolution has run its course, the
subsequent transformation from partons to hadrons should follow exactly
the same rules, whether in e$^+$e$^-$, ep or pp events. We have now seen
many LHC analyses that run counter to jet universality. The breaking notably
occurs at small transverse momenta, up to order 10~GeV, whereas jet
structure above that appears to follow expectations, as far as we can
tell. It is a fairly safe bet that the origin of universality breaking
is the close-packing of strings in the low-$p_{\perp}$ region.  The new
data has been very invigorating; the hadronization field, dormant for
many years, now is experiencing the emergence of new ideas. We have
highlighted some of them here --- QGP on one side, junctions, ropes and
shove on another --- but it remains to combine the latter ideas into a
consistent whole. And, as we have seen, the issues in soft QCD are not
limited to the hadronization itself, but is related to the preceding
multiparton interactions and colour reconnection descriptions as well,
e.g.\ to set the stage for close-packing effects to occur. 

In this article there has only been space to bring up a few of the data
and models being discussed today. Several more have been presented in other
contributions to this conference. It is clear that continued activity is
called for, not only for LHC itself but also to help prepare for future
colliders, such as EIC, ILC and FCC, and to strengthen ties to other fields
of study, such as fixed-target neutrino interactions and cosmic-ray
cascades.

\acknowledgments
Work supported by the Swedish Research Council, contract number 2016-05996.

%%\begin{thebibliography}{99}
\bibliographystyle{JHEP}
\bibliography{sjostrand.bib}

\providecommand{\href}[2]{#2}\begingroup\raggedright\begin{thebibliography}{10}

\bibitem{Bierlich:2022pfr}
C.~Bierlich et~al., \emph{{A comprehensive guide to the physics and usage of
  PYTHIA 8.3}},  \href{https://arxiv.org/abs/2203.11601}{{\ttfamily
  2203.11601}}.

\bibitem{Buckley:2011ms}
A.~Buckley et~al., \emph{{General-purpose event generators for LHC physics}},
  \href{https://doi.org/10.1016/j.physrep.2011.03.005}{\emph{Phys. Rept.}
  {\bfseries 504} (2011) 145}
  [\href{https://arxiv.org/abs/1101.2599}{{\ttfamily 1101.2599}}].

\bibitem{Bellm:2019zci}
J.~Bellm et~al., \emph{{Herwig 7.2 release note}},
  \href{https://doi.org/10.1140/epjc/s10052-020-8011-x}{\emph{Eur. Phys. J. C}
  {\bfseries 80} (2020) 452}
  [\href{https://arxiv.org/abs/1912.06509}{{\ttfamily 1912.06509}}].

\bibitem{Sherpa:2019gpd}
{\scshape Sherpa} collaboration, \emph{{Event Generation with Sherpa 2.2}},
  \href{https://doi.org/10.21468/SciPostPhys.7.3.034}{\emph{SciPost Phys.}
  {\bfseries 7} (2019) 034} [\href{https://arxiv.org/abs/1905.09127}{{\ttfamily
  1905.09127}}].

\bibitem{Alwall:2014hca}
J.~Alwall, R.~Frederix, S.~Frixione, V.~Hirschi, F.~Maltoni, O.~Mattelaer
  et~al., \emph{{The automated computation of tree-level and next-to-leading
  order differential cross sections, and their matching to parton shower
  simulations}}, \href{https://doi.org/10.1007/JHEP07(2014)079}{\emph{JHEP}
  {\bfseries 07} (2014) 079} [\href{https://arxiv.org/abs/1405.0301}{{\ttfamily
  1405.0301}}].

\bibitem{Alioli:2010xd}
S.~Alioli, P.~Nason, C.~Oleari and E.~Re, \emph{{A general framework for
  implementing NLO calculations in shower Monte Carlo programs: the POWHEG
  BOX}}, \href{https://doi.org/10.1007/JHEP06(2010)043}{\emph{JHEP} {\bfseries
  06} (2010) 043} [\href{https://arxiv.org/abs/1002.2581}{{\ttfamily
  1002.2581}}].

\bibitem{Campbell:2022qmc}
J.M.~Campbell et~al., \emph{{Event Generators for High-Energy Physics
  Experiments}},  in \emph{{2022 Snowmass Summer Study}}, 3, 2022
  [\href{https://arxiv.org/abs/2203.11110}{{\ttfamily 2203.11110}}].

\bibitem{Pierog:2013ria}
T.~Pierog, I.~Karpenko, J.M.~Katzy, E.~Yatsenko and K.~Werner, \emph{{EPOS LHC:
  Test of collective hadronization with data measured at the CERN Large Hadron
  Collider}}, \href{https://doi.org/10.1103/PhysRevC.92.034906}{\emph{Phys.
  Rev. C} {\bfseries 92} (2015) 034906}
  [\href{https://arxiv.org/abs/1306.0121}{{\ttfamily 1306.0121}}].

\bibitem{Bartalini:2018qje}
P.~Bartalini and J.R.~Gaunt, eds., \emph{{Multiple Parton Interactions at the
  LHC}}, vol.~29, WSP (2019),
  \href{https://doi.org/10.1142/10646}{10.1142/10646}.

\bibitem{CMS:2021lxi}
{\scshape CMS} collaboration, \emph{{Measurement of double-parton scattering in
  inclusive production of four jets with low transverse momentum in
  proton-proton collisions at $ \sqrt{s} $ = 13 TeV}},
  \href{https://doi.org/10.1007/JHEP01(2022)177}{\emph{JHEP} {\bfseries 01}
  (2022) 177} [\href{https://arxiv.org/abs/2109.13822}{{\ttfamily
  2109.13822}}].

\bibitem{ALEPH:2006bhb}
{\scshape ALEPH, DELPHI, L3, OPAL, LEP Electroweak Working Group}
  collaboration, \emph{{A Combination of preliminary electroweak measurements
  and constraints on the standard model}},
  \href{https://arxiv.org/abs/hep-ex/0612034}{{\ttfamily hep-ex/0612034}}.

\bibitem{Christiansen:2015yqa}
J.R.~Christiansen and P.Z.~Skands, \emph{{String Formation Beyond Leading
  Colour}}, \href{https://doi.org/10.1007/JHEP08(2015)003}{\emph{JHEP}
  {\bfseries 08} (2015) 003}
  [\href{https://arxiv.org/abs/1505.01681}{{\ttfamily 1505.01681}}].

\bibitem{Gieseke:2017clv}
S.~Gieseke, P.~Kirchgae\ss{}er and S.~Pl\"atzer, \emph{{Baryon production from
  cluster hadronisation}},
  \href{https://doi.org/10.1140/epjc/s10052-018-5585-7}{\emph{Eur. Phys. J. C}
  {\bfseries 78} (2018) 99} [\href{https://arxiv.org/abs/1710.10906}{{\ttfamily
  1710.10906}}].

\bibitem{ALICE:2021dhb}
{\scshape ALICE} collaboration, \emph{{Charm-quark fragmentation fractions and
  production cross section at midrapidity in pp collisions at the LHC}},
  \href{https://doi.org/10.1103/PhysRevD.105.L011103}{\emph{Phys. Rev. D}
  {\bfseries 105} (2022) L011103}
  [\href{https://arxiv.org/abs/2105.06335}{{\ttfamily 2105.06335}}].

\bibitem{ALICE:2021rzj}
{\scshape ALICE} collaboration, \emph{{Measurement of Prompt D$^{0}$,
  $\Lambda_{c}^{+}$, and $\Sigma_{c}^{0,++}$(2455) Production in
  Proton\textendash{}Proton Collisions at $\sqrt s$ = 13\,\,TeV}},
  \href{https://doi.org/10.1103/PhysRevLett.128.012001}{\emph{Phys. Rev. Lett.}
  {\bfseries 128} (2022) 012001}
  [\href{https://arxiv.org/abs/2106.08278}{{\ttfamily 2106.08278}}].

\bibitem{LHCb:2019fns}
{\scshape LHCb} collaboration, \emph{{Measurement of $b$ hadron fractions in 13
  TeV $pp$ collisions}},
  \href{https://doi.org/10.1103/PhysRevD.100.031102}{\emph{Phys. Rev. D}
  {\bfseries 100} (2019) 031102}
  [\href{https://arxiv.org/abs/1902.06794}{{\ttfamily 1902.06794}}].

\bibitem{Norrbin:2000zc}
E.~Norrbin and T.~Sj{\"o}strand, \emph{{Production and hadronization of heavy
  quarks}}, \href{https://doi.org/10.1007/s100520000460}{\emph{Eur. Phys. J. C}
  {\bfseries 17} (2000) 137}
  [\href{https://arxiv.org/abs/hep-ph/0005110}{{\ttfamily hep-ph/0005110}}].

\bibitem{LHCb:2021xyh}
{\scshape LHCb} collaboration, \emph{{Observation of a
  $\Lambda_b^0-\overline{\Lambda}_b^0$ production asymmetry in proton-proton
  collisions at $\sqrt{s} = 7 \textrm{ and } 8\,\textrm{TeV}$}},
  \href{https://doi.org/10.1007/JHEP10(2021)060}{\emph{JHEP} {\bfseries 10}
  (2021) 060} [\href{https://arxiv.org/abs/2107.09593}{{\ttfamily
  2107.09593}}].

\bibitem{ALICE:2016fzo}
{\scshape ALICE} collaboration, \emph{{Enhanced production of multi-strange
  hadrons in high-multiplicity proton-proton collisions}},
  \href{https://doi.org/10.1038/nphys4111}{\emph{Nature Phys.} {\bfseries 13}
  (2017) 535} [\href{https://arxiv.org/abs/1606.07424}{{\ttfamily
  1606.07424}}].

\bibitem{Bierlich:2014xba}
C.~Bierlich, G.~Gustafson, L.~L{\"o}nnblad and A.~Tarasov, \emph{{Effects of
  Overlapping Strings in pp Collisions}},
  \href{https://doi.org/10.1007/JHEP03(2015)148}{\emph{JHEP} {\bfseries 03}
  (2015) 148} [\href{https://arxiv.org/abs/1412.6259}{{\ttfamily 1412.6259}}].

\bibitem{Bierlich:2016vgw}
C.~Bierlich, G.~Gustafson and L.~L{\"o}nnblad, \emph{{A shoving model for
  collectivity in hadronic collisions}},
  \href{https://arxiv.org/abs/1612.05132}{{\ttfamily 1612.05132}}.

\end{thebibliography}\endgroup
%%\end{thebibliography}

\end{document}